\documentclass{article}
\usepackage{spconf,amsmath,graphicx,refstyle,amsfonts}

\usepackage[T1]{fontenc} 
\usepackage[utf8]{inputenc} 
\usepackage{url}
\usepackage{color}
\usepackage[font=footnotesize]{subfig}
\graphicspath{{./figs/}}

\renewcommand{\vec}[1]{\mathbf{#1}}

\newcommand{\mM}{\vec{M}}
\newcommand{\mY}{\vec{Y}}
\newcommand{\mX}{\vec{X}}
\newcommand{\mO}{\vec{O}}

\newcommand{\vy}{\vec{y}}

\title{Learned Complex Masks for Multi-Instrument Source Separation}
\name{Andreas Jansson$^{2}$, Rachel M. Bittner$^1$, Nicola Montecchio$^1$, Tillman Weyde$^2$}
\address{$^1$ Spotify, $^2$ City University of London}
\begin{document}
%
\maketitle
\begin{abstract}

Music source separation in the \mbox{time-frequency} domain is commonly
achieved by applying a soft or binary mask to the magnitude component
of (complex) spectrograms.  The phase component is usually not
estimated, but instead copied from the mixture and applied to the
magnitudes of the estimated isolated sources.  While this method has
several practical advantages, it imposes an upper bound on the
performance of the system, where the estimated isolated sources inherently exhibit audible ``phase artifacts''.  In this paper we
address these shortcomings by directly estimating masks in the complex
domain, extending recent work from the speech enhancement literature.  The method is particularly well suited for multi-instrument musical source separation since residual phase artifacts are more pronounced for spectrally overlapping instrument sources, a common scenario in music.  
We show that complex masks result in better separation than masks that operate solely on the magnitude component.
\end{abstract}
\section{Introduction}\label{sec:introduction}

Musical audio mixtures can be modeled as a sum of individual sources, and musical audio source separation aims to estimate these individual sources given the mixture.
A common way to frame this problem is to separate a music recording into percussion, bass, guitar, voice, and ``other'' signals.
\emph{Spectral masking} is a common technique used in this context:
a ``mask'' of the signal's spectrogram is estimated, whose aim is to selectively ``let through'' only the components of the input signal corresponding to the particular source \cite{wang2005ideal}.


Typically, masks (both binary and soft) are only applied to the
magnitude component of the input (complex) spectrogram, and the phase from the mixture is reused for each of the sources.
This results in significant phase artifacts: sources that overlap in time and frequency with the target source leave audible traces in the masked signal.
This effect is more pronounced the greater the spectral overlap between sources is, and is especially problematic in the music domain: due to the very nature of music, musical instruments and voices are often synchronous in time by following a shared meter (unlike speech), and their spectral components tend to occupy the same frequency ranges.

In this paper we present an approach 
that uses
\emph{complex-valued} masks as a drop-in replacement for \mbox{magnitude-domain} masks.

\begin{figure*}[h!]
  \centering
  \includegraphics[width=0.85\textwidth]{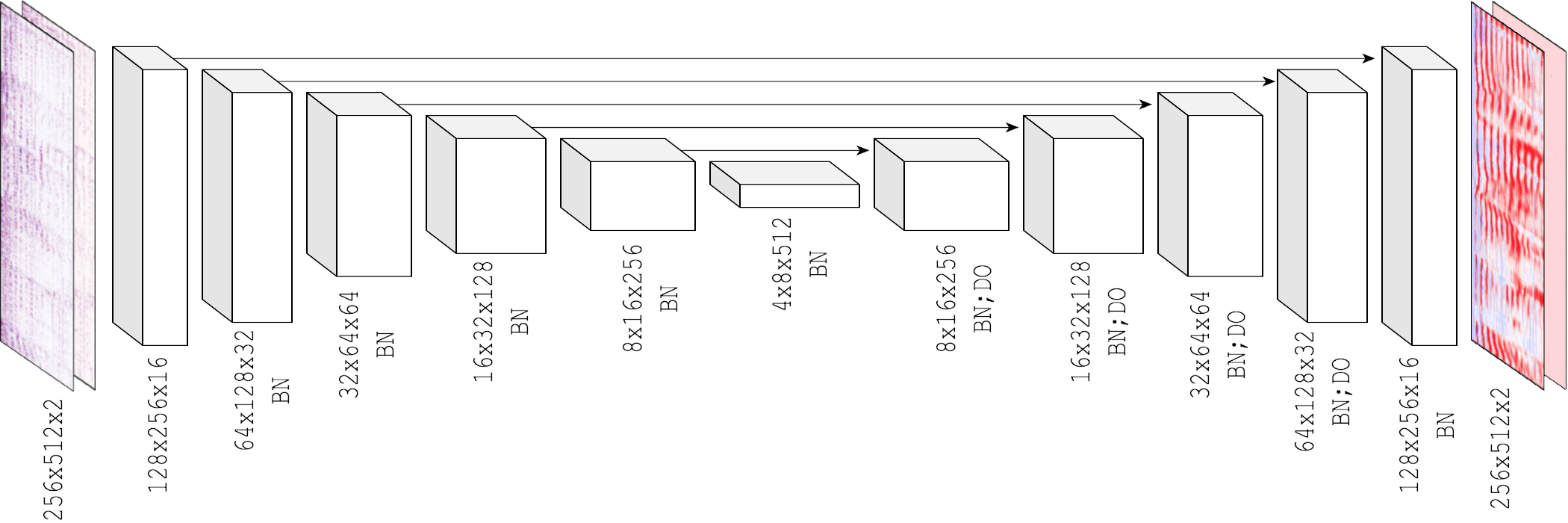}
  \caption{Architecture diagram of the proposed model.}
  \label{fig:architecture-diagram}
\end{figure*}

\section{Related Work}

Neural network-based approaches have been applied to the problem of
audio source separation for over a decade.
For a review of such techniques in the speech domain, see \cite{wang2018supervisedspeech}.

In the music domain, the majority of Neural network-based models operates on
time-frequency representations (spectrograms) of both the input and
output signals; usually separation is achieved using masks in the
magnitude domain.
Different network architectures -- both convolutional \cite{deepkaraoke, jansson2017singing} and recurrent \cite{huang2014deep} -- have been used to separate vocals from accompaniment, as well as for multi-instrument separation.  
Skip connections, introduced
in \cite{resnet}, have emerged as a common feature of source
separation networks, because of their ability to carry fine-grained
details from the input mixture representation to the layers closer to
the final output representation.
%
as it makes the training procedure more efficient.
The U-Net \cite{originalunet} 
is a relatively simple application of skip connections: an
\mbox{hourglass-shaped} convolutional \mbox{encoder-decoder} pair is
augmented with connections between earlier layers of the encoder to
later layers in the decoder.  U-Nets have been
applied to musical source separation in
\cite{jansson2017singing,park2018musicsource}.  

A number of recent systems propose alternative models
that avoid dealing with phase information altogether,
by operating directly in the time domain:
in \cite{rethage2018wavenetspeechdenoising}, the authors construct a
non-causal WaveNet \cite{oord2016wavenet} for speech enhancement
entirely in the time domain; \mbox{Wave-U-Net}
\cite{stoller2018wave} is an extension of WaveNet that adds skip
connections in the time domain, analogous to the U-Net in the
frequency domain.

Other approaches attempt to explicitly 
make use of phase in the \mbox{time-frequency} domain:
in \cite{erdogan2015phase} incorporating phase
information in the mask function results in more accurate source
estimation, even though the mask itself is real-valued.  
Predicting phase directly appears to be intractable
\cite{williamson2016complexratio}, presumably because of the lack of
inherent structure due to the wrap-around nature of phase;
proposed solutions make use of \emph{phase unwrapping} \cite{mayer2017impact,spoorthi2019phasenet}
or discretization of phase \cite{takahashi2017multimulti}.

In the speech domain, the complex Ideal Ratio Mask (cIRM) was
introduced in \cite{williamson2016complexratio} as the real and
imaginary ratios between complex source and mixture spectrograms.
The Phase-U-Net \cite{choi2019phasespeech} estimates complex masks, either using a
real-valued internal representation or introducing the complex-valued
equivalent of the traditional neural network building blocks such as
matrix multiplication and convolution; the paper also deals with
masking using a complex representation based on polar coordinates, and
constraining the magnitude component. 

Typically, methods based on a time-frequency representation are trained
using a loss function which measures the distance (in L1 or L2 norm)
between a target spectrogram (coming from an \mbox{individually-recorded}
instrument, subsequently mixed to create the input signal)
and the input spectrogram after the application of the predicted mask.
More sophisticated designs allow a more direct optimization of  evaluation measures, such as \mbox{source-to-distortion} ratio (SDR)
\cite{venkataramani2018end, choi2019phasespeech}.



\section{Model}

\subsection{Datasets}

The proposed model is trained on a dataset of about 2000 professional-quality recordings.  For each recording, five
\emph{stems} are available, grouped by instrument: ``Vocals'', ``Guitar'',
``Bass'', ``Percussion'', and ``Other''.  Each recording is, by
design, the (unweighted) sum of its five stems.

To increase the size of the training set, a series of augmentation
steps is performed on the stems.  The augmentations include global,
\mbox{track-level} changes -- such as a time stretching, pitch
shifting, or resampling -- applied across all stems, and individual,
\mbox{stem-level} changes, such as volume adjustments,
equalization/filtering, effects such as phasers and flangers, and reverb.
For each example in the training set, 10 random augmentations are
generated.

For validation and testing, the MUSDB 2018 dataset~\cite{stoter2018}
is used. 
MUSDB contains only 4 instrument types, namely ``Vocals'', ``Bass'', ``Percussion'', and ``Other''.
When running validation and test results, we group our model's outputs for ``Guitar'' into the ``Other'' category.

\subsection{Architecture}

The proposed model extends the U-Net source separation model detailed
in \cite{jansson2017singing}, which in turn extends
\cite{isola2017imageimage}; pictured in
Figure~\ref{fig:architecture-diagram}, it consists of 6 downsampling
``encoder'' layers, 6 upsampling ``decoder'' layers, and skip
connections between corresponding encoder and decoder layers.  
The network operates on spectrograms extracted from 22050 Hz mono audio signals, with
overlapping windows of size 1024, and hop 
size 256.
Each encoder layer comprises a strided convolution (with kernel size
$5\times5$ and stride $2\times2$) followed by batch normalization
(except in the first layer), and a leaky ReLU activation function.  In
the decoder, each layer consists of a strided transpose convolution (with
kernel size $5\times5$ and stride $2\times2$, as in the encoder);
batch normalization is applied to all decoder layers except the last,
as well as 50\% dropout in the first five layers, and a ReLU
activation function in all layers except the last.

\subsection{Input and Output Representation}

For multi-source separation, independent models are trained for
vocals, drums, guitar, and bass.  The ``Other'' source is estimated as
the residual difference between mixture and the sum of the four
estimated sources.  Each \mbox{source-specific} network is presented with
mixture spectrograms, which during training are sliced into patches of 256 frames,
and fed in batches of 16 patches.  This is not necessary
during inference, since the network is fully convolutional and can be
applied to the spectrogram of the full length of the signal.

The network is trained to output a mask that is then applied to the
mixture spectrogram to produce the source spectrogram estimate.  We
evaluate two variations of this architecture: non-complex, where the
input is a single mono channel representing the magnitude of the
mixture spectrogram; and complex, where the input is the complex
spectrogram, represented as independent real and imaginary channels.
In the non-complex case, the mask is only applied to the magnitude
component of the mixture, whereas in the complex case the
mask is applied in the complex domain.  After mask application, we synthesize the
estimated source signal by means of the Inverse Short-Time Fourier
Transform (ISTFT).

It should be noted that in the complex case, although the input and
output representations are both complex, the network itself is
internally real-valued.

\subsection{Masking}

In the remainder of this section, $\mO^{\mathbb{C}}$ and $\mO^{\mathbb{R}}$ denotes the outputs by the complex and non-complex networks; $\mM^{\mathbb{C}}$ and $\mM^{\mathbb{R}}$ denotes complex and non-complex masks; $\mX$ denotes the spectrogram of the input mixture to be separated by the model; for individual stems (individual instruments or voice), $\mY$ and $\hat{\mY}$ denote the spectrograms of the recorded and estimated stems, respectively, while $\vy$ and $\hat{\vy}$ denote their time-domain equivalents.

\subsubsection{Non-complex masking}
\label{lab:non-complex-masking}

The non-complex mask is computed as $\mM^{\mathbb{R}} = \sigma \left( \mO^{\mathbb{R}}\right)$ where $\sigma$ is the sigmoid function, constraining the (real-valued) network output to the $(0, 1)$ range.
It is multiplied element-wise (denoted by $\otimes$) with the mixture spectrogram magnitude to obtain the estimated source spectrogram magnitude.
The phase is copied from the mixture spectrogram into the spectrogram of the estimated stem, unaltered.

\begin{equation}
\hat{\mY} = \mM^{\mathbb{R}} \otimes \lvert \mX \rvert \otimes e^{i \angle \mX}
\end{equation}

\begin{figure*}[h!]
  \centering
  \includegraphics[width=0.9\textwidth]{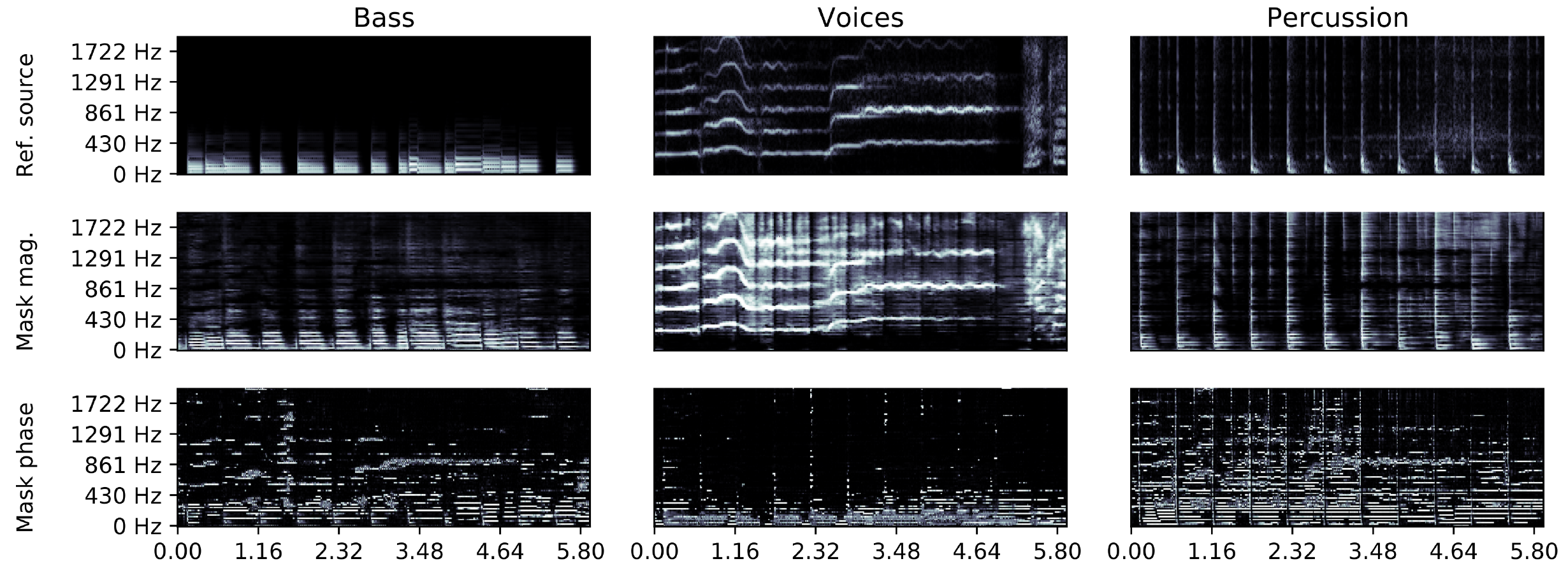}
  \caption{Magnitude and phase of learned complex masks (Complex-SDR) for Bass, Voices, and Percussion on a test example from MUSDB18, and the reference STFTs for each source. \textbf{Top Row:} Magnitude (dB) of the ideal STFTs. \textbf{Middle Row:} Magnitude of the learned complex masks.  \textbf{Bottom Row:} Absolute value of the phase of the learned complex masks.}
  \label{fig:mask-plot}
\end{figure*}

\subsubsection{Complex masking}
\label{lab:complex-masking}

In the complex case we estimate a complex-domain mask as in \cite{choi2019phasespeech}, constrained in 
magnitude but not in phase:

\begin{align}
    |\mM^{\mathbb{C}}| & = \tanh ( |\mO^{\mathbb{C}}| ) \\
    \angle \mM^{\mathbb{C}} &= \mO^{\mathbb{C}} \oslash |\mO^{\mathbb{C}}|
\end{align}

where $\oslash$ denotes element-wise division.
The complex mask is then applied \mbox{element-wise}, as:
\begin{equation}
\hat{\mY} = \lvert\mM^{\mathbb{C}}\lvert \otimes \lvert \mX \rvert \otimes e^{i (\angle \mM^{\mathbb{C}} + \angle \mX)}
\end{equation}

The mask applied determines a magnitude rescaling of the mixture STFT and a rotation (addition) in phase.
This allows the model not only to remove interfering frequency bins, but also to cancel the phase from other sources.

\subsection{Loss}

We evaluate three different loss functions: \emph{Magnitude loss}, \emph{SDR loss}, and the combination \emph{SDR + Magnitude loss}.

The \emph{Magnitude loss} is optimized by minimizing the L1 distance between the spectrogram magnitudes of the estimated and 
target components:
\begin{equation}
\mathcal{L}_{\textrm{Mag}} = \lvert \mY - \hat{\mY} \rvert_1
\end{equation}

\noindent The time-domain \emph{SDR loss} \cite{choi2019phasespeech} is defined as:

\begin{equation}
\mathcal{L}_{\textrm{SDR}} = - \frac{\vy \cdot \hat{\vy}}{\Vert \vy \rVert \Vert \hat{\vy} \rVert}
\end{equation}
which smoothly upper bounds the SDR metric we are interested in optimizing.

\noindent The hybrid time-domain and frequency-domain \emph{SDR + Magnitude loss} is defined as:
\begin{equation}
\mathcal{L}_{\textrm{SDR+Mag}} = \mathcal{L}_{\textrm{SDR}} +  \mathcal{L}_{\textrm{Mag}}
\end{equation}


\section{Results}

\begin{figure} 
  \centering
        \includegraphics[width=0.9\columnwidth]{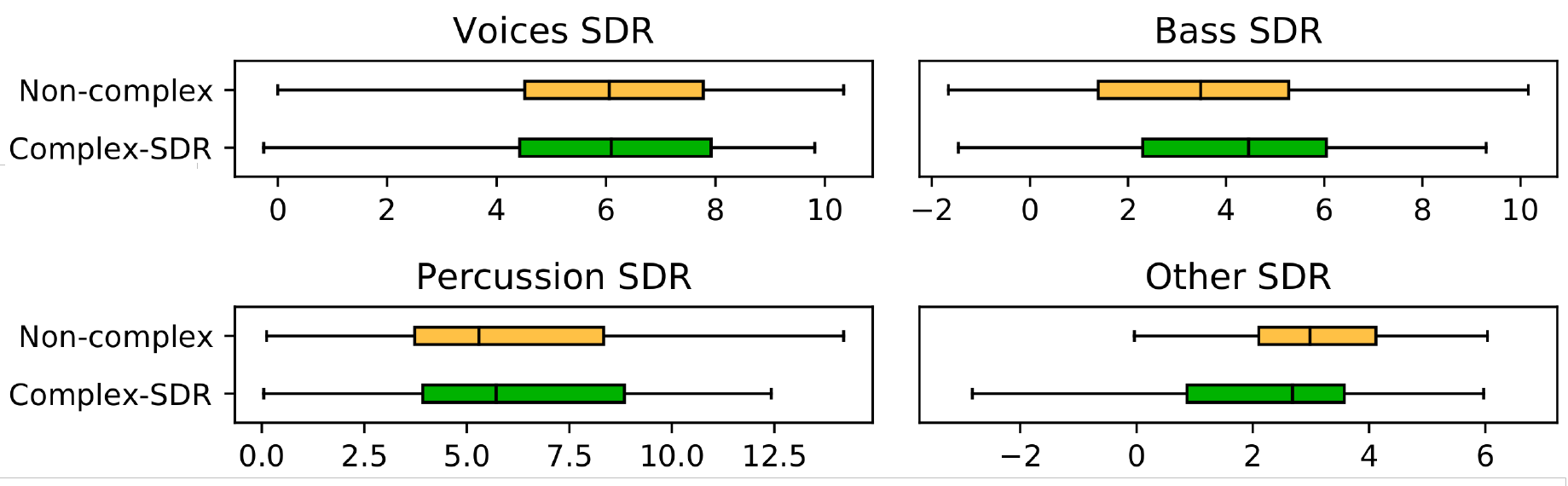}
        \caption{Signal to Distortion Ratio 
        across experiments.}
        \label{fig:bss-metrics}
        
\end{figure}

Two configurations of the system (``Non-complex'', ``Complex-SDR'') are evaluated on the MUSDB 18 dataset, using the \texttt{museval} toolkit \cite{SiSEC18}; results are shown in Figure~\ref{fig:bss-metrics}.

There are slight differences between the results of the different models when considering different metrics, 
however the complex model achieves higher Signal to Distortion Ratio (commonly considered the most important metric) than the non-complex model for most sources.
Although these results do not compare favorably to \mbox{State-of-the-Art} approaches, such as those
reported in the yearly SiSEC evaluation campaign~\cite{SiSEC18},
they do yield insights into the design of such systems, by providing a unified comparison not only in terms of dataset, but also through the use of a single model in which individual aspects (masking, loss function) are controlled.

We also collected subjective judgments from a number of individuals, using a methodology similar to that of \cite{cartwright2018crowdsourced}.
A questionnaire was prepared, in which subjects were prompted to compare audio recordings in pairs, each pair presenting the outputs of two different model configurations applied the same short audio excerpt.
As a reference, evaluators were given both the original mixture and the target single source signal.

Two questions were asked per pair of audio excerpts: ``Quality: Which one has better sound quality?'', and ``Separation: Which one has better isolation from the other instruments in the original mix?''.
We collected 210 judgements from 8 individuals trained in audio engineering.

We found that in 58\% of the ``Separation'' questions, SDR loss was preferred to non-complex loss.
The effect was stronger for the ``Quality'' question, where 65\% of judgments preferred SDR loss.
The preference for SDR loss was most pronounced on bass, followed by percussion and vocals.





\subsection{Qualitative Analysis}

A website with audio examples of the complex and non-complex masks can be found online~\footnote{ \url{http://bit.ly/lcmmiss }}.
The biggest audible difference between the audio output by the complex masking model and the non-complex masking model is the amount of phase distortion, in particular for bass and percussion.
When using real masks, both bass and percussion tend to sound very tinny, and the phase of other sources is often captured in the sound of the reconstructed source, such as hearing faint voices in a percussion signal.
The predicted vocals sound similar in both models, and the overall quality is rather good.

\subsection{Analysis of learned masks}
Figure~\ref{fig:mask-plot} displays the learned masks from the Complex-SDR model on a recording, and the STFTs of the clean reference sources;
the middle and bottom rows show the magnitude and (absolute) phase of the learned masks respectively.
As expected, the magnitude of the learned masks looks like that of a typical soft mask, where the time frequency bins of the target source have values close to 1 and other bins have values close to 0.
For bass, harmonic patterns with few transients in the low frequency range are noticeable; for voice, harmonic patterns tend to inhabit a higher frequency range and exhibit stronger transients, e.g., at the starts of words; percussion exhibits more regular transients.

The phase of the mask determines how much the phase of the mixture should be rotated:
higher values indicate places where the phase of the mixture is substantially different from the phase of the source (e.g., because it overlaps with another source).
In this example, we see that for percussion there are large phase corrections being made for horizontal patterns, likely canceling the phase from harmonic sources in the mixture.
Additionally, the phase is corrected the instant before an onset, which makes the onset sound more crisp than if it were smeared with phase from other sources.
For the vocal signal, most of the phase corrections are made where the bass and percussion masks are active, indicating that the phase of the bass and percussion is being canceled out for the vocal mask.
Close inspection of the phase of the bass mask reveals that lower frequency bands that have large values from the mask and for the phase are interleaved, implying that the phase of closely overlapping low frequency sources is being canceled.

\section{Conclusions}

The use of complex masks for multi-instrument source separation is explored using a U-Net architecture.
It was  found that learned complex masks perform better than non-complex masks perceptually, and marginally better in terms of SDR.
Analysis of the learned masks highlights a particularly strong effect on percussion and bass.
As a result, the constrained complex mask is a practical, \mbox{low effort}, drop-in replacement for common magnitude-domain masks.

\bibliographystyle{IEEEbib}
\bibliography{references}

\end{document}